# Thermodynamical Scaling of the Glass Transition Dynamics

R.Casalini[1,2,*] and C.M.Roland[1†]

[1]Naval Research Laboratory, Code 6120, Washington DC 20375-5342
[2]George Mason University, Fairfax, Virginia 22030

Electronic address: [*]casalini@.ccs.nrl.navy.mil [†]roland@nrl.navy.mil

**Abstract**

Classification of glass-forming liquids based on the dramatic change in their properties upon approach to the glassy state is appealing, since this is the most conspicuous and often-studied aspect of the glass transition. Herein, we show that a generalized scaling, $\log(\tau) \propto T^{-1}V^{-\gamma}$ , where $\gamma$ is a material-constant, yields superpositioning for ten glass-formers, encompassing van der Waals molecules, associated liquids, and polymers. The exponent $\gamma$ reflects the degree to which volume, rather than thermal energy, governs the temperature and pressure dependence of the relaxation times.

Studies of the glass transition have been ongoing for many years, due to both the fundamental and the practical significance of the phenomenon. Theoretical efforts remain at the model building stage, with no consensus reached concerning the detailed causes of the slowing down and ultimate arrest of molecular motions.[1-7] It is generally agreed that the spectacular loss of molecular mobility upon approach to the glassy state reflects (i) increased molecular packing and consequent jamming, as well as (ii) a decrease in thermal energy which causes trapping of molecules in the potential wells of the energy landscape. The relative importance of these two factors, volume and thermal energy, is a contentious issue.[8-13] It cannot be resolved from the usual practice of employing only temperature variations in experimental studies, since the volume and thermal energy both depend on temperature. The problem is circumvented by using pressure, $P$, as an experimental variable,[14, 15] whereby the volume, $V$, can be altered while maintaining temperature, $T$, constant. From recent experiments of this type, it has become clear that for most glass formers, neither $T$ nor $V$ is the dominant variable controlling the temperature-dependence of the relaxation times. For van der Waals liquids, $T$ and $V$ play an almost equivalent role, while for polymers $T$ assumes a somewhat greater importance, although volume effects are not negligible.[8, 10-13] Only for strongly hydrogen-bonded liquids $T$ does appear to become dominant.[9, 16] These results indicate that neither $T$ or $V$ is the appropriate thermodynamic variable for describing the dynamics of glass-formers, but rather some function of these two quantities is required to uniquely represent relaxation times, viscosities, etc.

Tölle et al.[17] deduced from inelastic neutron scattering under high pressure that the behavior of *o*-terphenyl (OTP) can be modeled as soft spheres interacting with an $r^{-12}$ repulsive potential. This hypothesis yields a description of OTP in terms of a single quantity $T^{-1}V^{-4}$. Subsequently, Dreyfus et al.[12] showed that viscosity data for OTP can be effectively rescaled onto a master curve when plotted as a function of $T^{-1}V^{-4}$. Since we recently measured dielectric relaxation times over a wide range of frequencies, temperatures, and pressures for glass-formers for which we also determined the equation of state parameters, *V*(*T*,*P*), it is of interest to attempt to extend the scaling of Tölle[17] to these other materials. Our analysis reveals strong deviations from the proposed $T^{-1}V^{-4}$ scaling. For example, in Fig.1 dielectric relaxation times measured for 1,1'-di(4-methoxy-5-methylphenyl)cyclohexane (BMMPC)[11] and D-sorbitol[16] are displayed versus the $T^{-1}V^{-4}$. Clearly, the data do not superpose when plotted in this fashion.

The use of $T^{-1}V^{-4}$ follows directly from an assumed $r^{-12}$ repulsive potential; however, this is just a special case of the more general $r^{-n}$ interaction energy.[18] Accordingly, the scaling in Fig. 1 can be considered as one specific application of a more general relationship, $\log(\tau) \propto T^{-1}V^{-\gamma}$, with $\gamma$ a material-constant, that can be applied to all glass-formers.

In Fig.2 the dielectric relaxation times are shown for four materials ((a) BMMPC[11], (b) phenolphthalein-dimethylether (PDE)[10], (c) D-sorbitol[16], and (d) 1,2 polybutadiene (1,2-PB)[19]) measured at varying *T* at atmospheric pressure (0.1MPa) and varying *P* at constant temperature (as listed in the figure). These are representatives of various classes of prototypical glass-formers: strong van der Waals liquids, moderately fragile van der Waals liquids, H-bonded liquids and polymers. The data are plotted versus $T^{-1}V^{-\gamma}$, where $\gamma$, which equals 4 in the model of Tölle, is adjusted to yield superpositioning of the curves. It can be seen that for a given, material-specific value of $\gamma$ (as reported in the figure), a master curve is obtained, encompassing nine decades of frequency and very different conditions of $T$ and $V$. This scaling is able to accurately depict the dynamics for both van der Waals and associated liquids, as well as polymers.

The parameter $\gamma$ provides a measure of the relative importance of $V$ as opposed to $T$. For strictly activated dynamics, in which thermal energy dominates the behavior, $\gamma$=0, while for the hard sphere limit, $\gamma\rightarrow\infty$.[18, 20] Therefore, we expect $\gamma$ to correlate with the ratio of the activation enthalpy at constant volume $E_V$(=$R\left[\partial\log\left(\tau\right)/\partial\left(T^{-1}\right)\right]_V$) to that at constant pressure $E_P$(=$R\left[\partial\log\left(\tau\right)/\partial\left(T^{-1}\right)\right]_P$), $E_V/E_P$. This ratio assumes the value of 1 or 0, for temperature or volume dominated dynamics, respectively.[8] In Fig.2 the parameter $\gamma$ determined herein for ten glass-formers is plotted versus this ratio $E_V/E_P$. We include the ideal case of $T$-dominated dynamics. The strong inverse correlation of the scaling parameter $\gamma$ with $E_V/E_P$ is evident.

The results in Fig.3 can be rationalized by considering a simple $V$ and $T$ dependence of $\tau$, corresponding to an activated process having a volume-dependent activation energy,

$$\tau\left(T,V\right)=\tau_0\exp\left(\frac{C}{TV^{\gamma}}\right) \qquad (1)$$

where $\tau_0$ and $C$ are constants. This relation gives a linear dependence of log($\tau$) on $T^{-1}V^{-\gamma}$, although such is not actually observed in Fig.2. Notwithstanding, from eq.1 it is straightforward to show that

$$\left.\frac{E_V}{E_P}\right|_{T_g}=\frac{1}{1-\gamma T_g\alpha_P} \qquad (2)$$

where $\alpha_P$ is the expansion coefficient at constant pressure ($\alpha_P < 0$). Taking the product $T_g\alpha_P$ to be constant as a first approximation, we can obtain a satisfactory description of the observed behavior. The best fit to the data in Fig.3 (solid line) gives $T_g\alpha_P = 0.19 \pm 0.01$. Interestingly, for the ten glass formers considered in Fig.3, we find $T_g\alpha_P = 0.16 \pm 0.03$.

The scaling parameter $\gamma$ is larger for glass-formers in which $V$ is the more dominant control variable, such that scaling the relaxation times according to $T^{-1}V^{-\gamma}$ effectively "removes" the volume contribution to the super-Arrhenius behavior observed near $T_g$. Deviation from an Arrhenius temperature-dependence is usually quantified by the steepness index, $m = d\log(\tau)/d(T_g/T)\big|_{T=T_g}$.[5] The activation enthalpy $E_P$ differs only by a constant from the fragility. In Fig.4 we have constructed a plot, analogous to a conventional fragility (or cooperativity) plot[5], but using as the abscissa $T^{1}V^{-\gamma}$ normalized by its value for $\tau$=10s. (This is a convenient reference point, which avoids extrapolation to more usual definitions of the glass temperature, such as $\tau$ = 100s.) In Fig.4 for BMMPC[11], in which $V$ is known to be more dominant than $T$ ($E_V/E_P$ = 0.41), the non-linearity seen in a conventional Arrhenius plot is almost removed. On the other hand, for sorbitol[16], in which T is the dominant control variable ($E_V/E_P$ = 0.86), strong curvature is maintained despite the $T^{1}V^{-\gamma}$ scaling. BMPC[11] and salol[21] have the same $m$, so that their fragility plots are very similar; however, in Fig. 3 they exhibit quite different curvatures, reflecting different relative contributions of $V$ and $T$ ($E_V/E_P$ = 0.38 and 0.43, respectively).

In conclusion, we have shown that the dynamics of glass-forming liquids and polymers can be described over many decades of frequency and a wide range of temperatures and volumes, through the use of the scaling parameter $T^{-1}V^{-\gamma}$. The exponent $\gamma$ is dependent on the material, and found to account for the volume contribution to the dynamics. A consequent modification of the usual fragility plot reveals directly the relative contributions to the super-Arrhenius character from temperature and volume, potentially "redefining" our concept of fragility.

The authors acknowledge K.L.Ngai for very helpful discussions and the Office of Naval Research for financial support.

**Figure captions**

**Fig.1** Dielectric relaxation time for BMMPC[11] and D-sorbitol[16] plotted versus the quantity $T^{-1}V^{-4}$. The data were measured at atmospheric pressure and various temperatures and at constant temperature and various pressures (listed in figure). Specific volumes were obtained from PVT measurements.

**Fig.2** Dielectric relaxation time measured versus the parameter $T^{-1}V^{-\gamma}$ for (a) BMMPC[11], (b) PDE [10], (c) D-sorbitol[16], and (d) 1,2-PB[19]. The data were measured varying $T$ at atmospheric pressure (0.1MPa) and varying $P$ at constant temperature (as listed in the figure). Specific volumes were obtained from PVT measurements.

**Fig.3** Ratio of the activation enthalpy at constant volume $E_V$ and at constant pressure $E_P$ plotted versus the parameter $\gamma$. The solid line is the best fit to the data of Eq.2. B: D-sorbitol[16]; C: 1,2-polybutadiene[19]; D: poly(vinyl methyl ether)[19]; E: poly(phenyl glycidy ether)-co-formaldehyde[10]; F: ortho-terphenyl[8, 12]; G: phenolphthalein-dimethylether[10]; H: polymethylphenylsiloxane[22]; I: phenyl salicylate[19]; J: 1,1'-bis(p-methoxyphenyl)cyclohexane[11]; K: 1,1'-di(4-methoxy-5-methylphenyl)cyclohexane[11].

**Fig.4** Dielectric relaxation time versus $T^{-1}V^{-\gamma}$ normalized by the value for which $\tau$=10s.

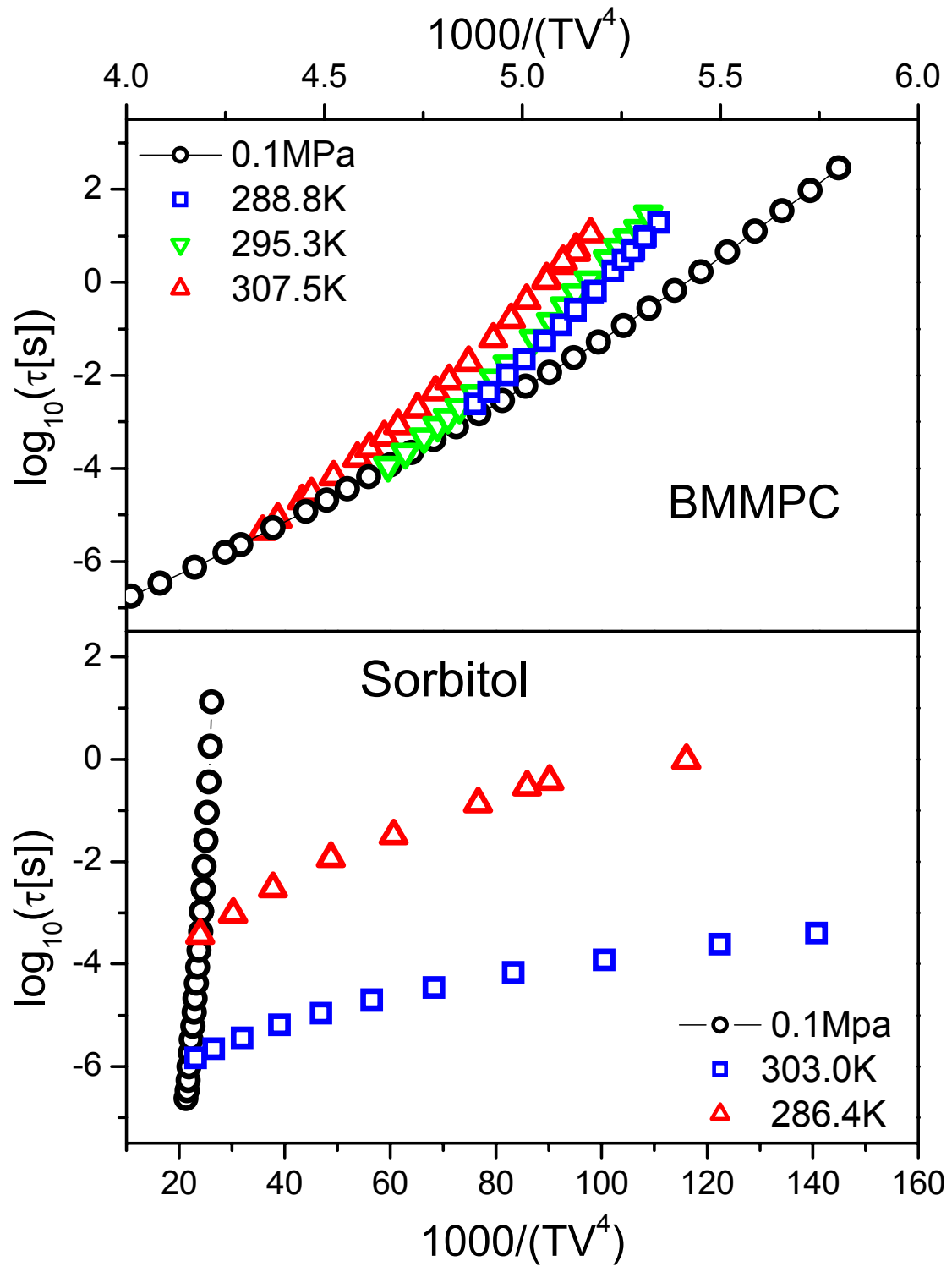

Figure 1 R.Casalini and C.M.Roland

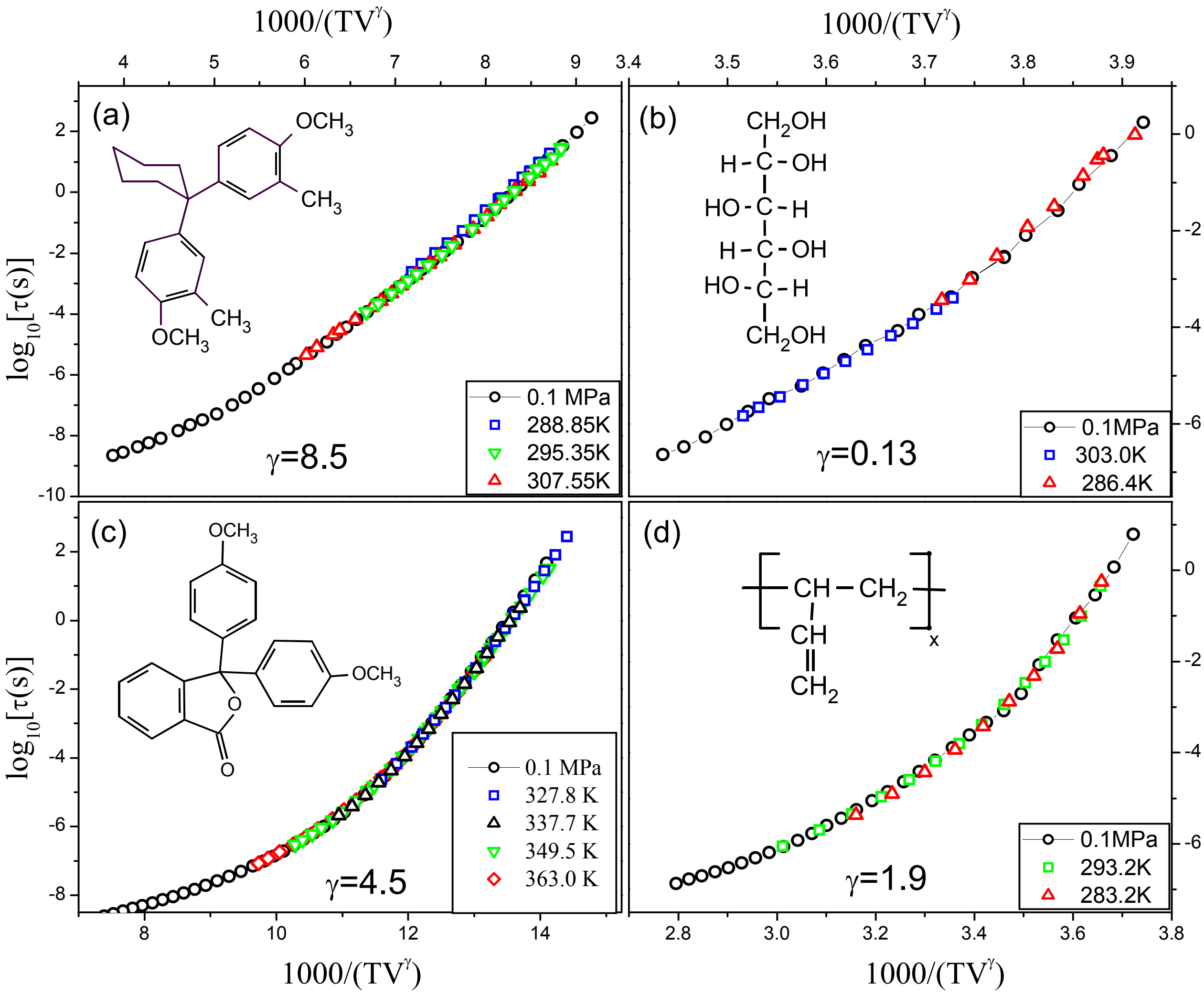

Figure 2 R.Casalini and C.M.Roland

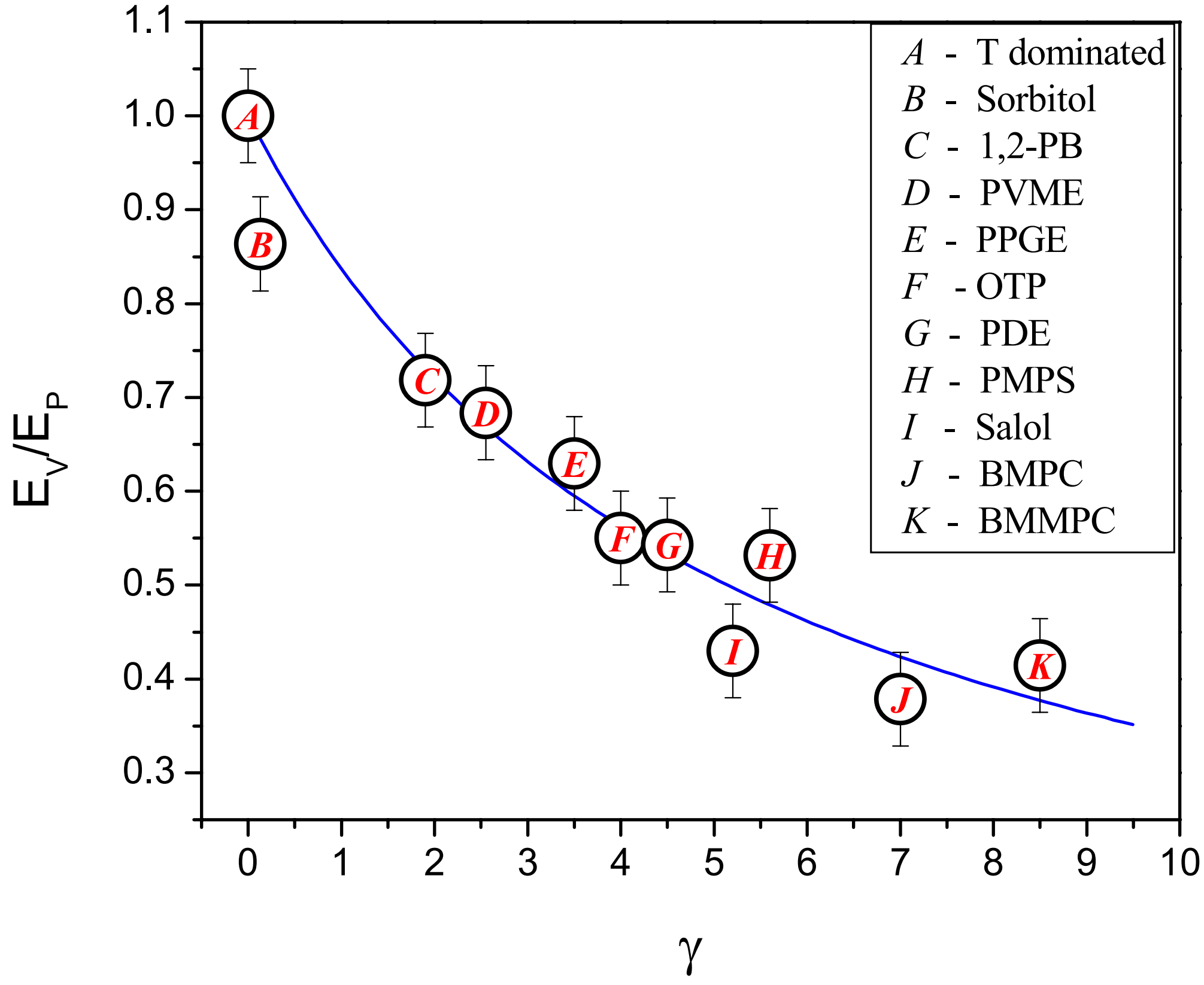

Figure 3 R.Casalini and C.M.Roland

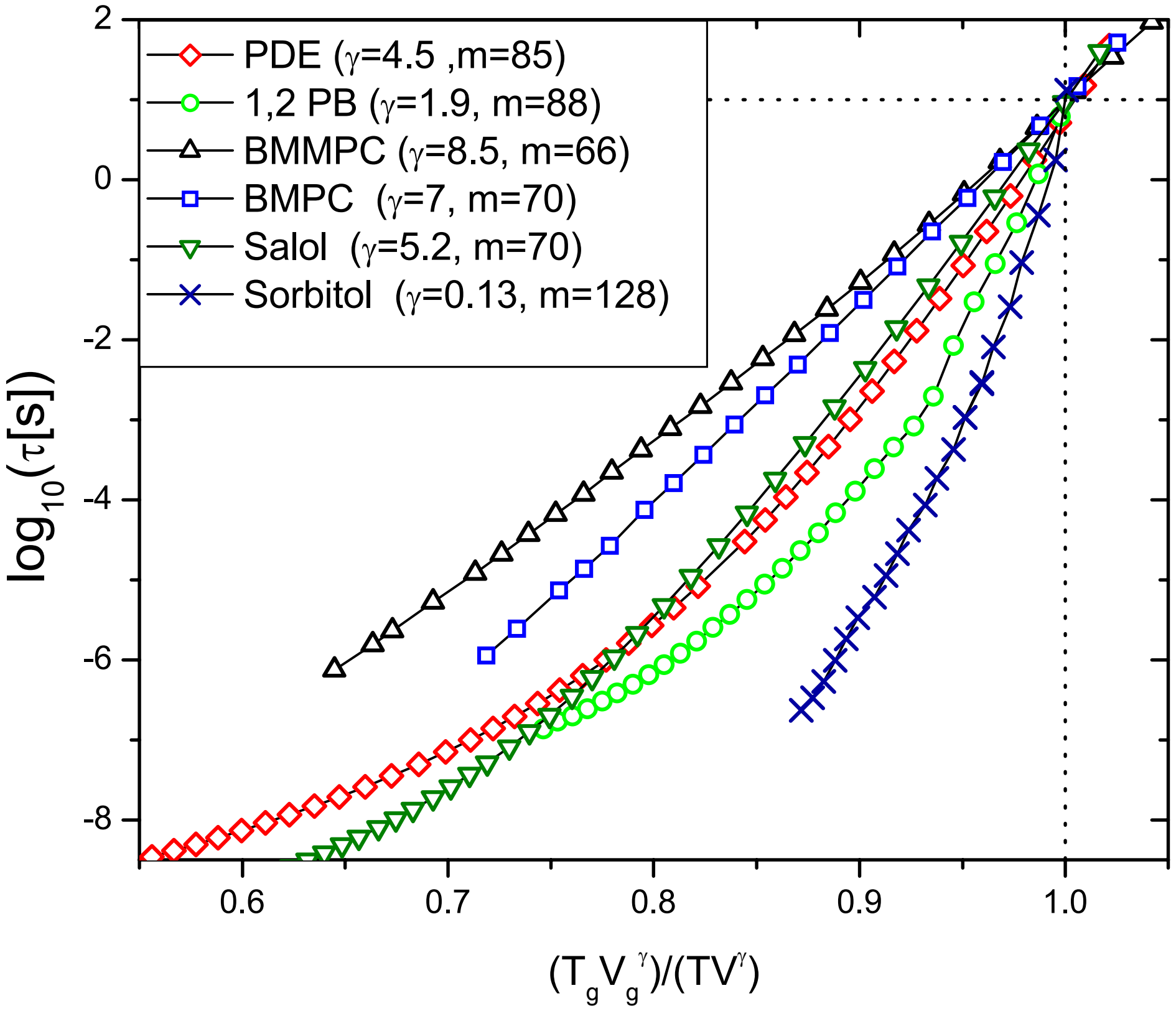

Figure 4 R.Casalini and C.M.Roland